\documentclass[11pt]{article}
\usepackage{amsmath,amsfonts,amsthm}
\newcommand{\R}{{\mathbb R}} \newcommand{\N}{{\mathbb N}}
 \newcommand{\Z}{{\mathbb Z}}
  \def\C{{\mathbb C}}
\newcommand{\Prm}{{\mathbb P}}

\newcommand{\E}{{\mathbb E\,}}
\renewcommand{\epsilon}{\varepsilon } 
\newcommand{\g}{\gamma } 
\renewcommand{\rho}{\varrho } 
 
\renewcommand{\phi}{\varphi }

\renewcommand{\b}{\beta }

\newcommand{\ran}{{\rm ran }}
\newcommand{\q}{{\rm q }}

\def\rs{\right>}
\def\lg{\left|}

\newtheorem{theorem}{Theorem}
\newtheorem{lemma}{Lemma}
\newtheorem{corollary}{Corollary}
\newtheorem{proposition}{Proposition}
\begin {document}
 \title{
 On the Power of Quantum Algorithms for Vector Valued Mean Computation}

\author {Stefan Heinrich\\
Dept. Computer Science\\
University of  Kaiserslautern\\
 Germany\\
e-mail: heinrich@informatik.uni-kl.de\\
homepage:\\http://www.uni-kl.de/AG-Heinrich}

\date{}
\maketitle 
\begin{abstract}
We study computation of the mean of sequences with values in finite dimensional 
normed spaces and compare the computational
power of classical randomized with that of quantum algorithms for this problem. It turns out that 
in contrast to the known superiority of quantum algorithms in the scalar case, 
in high dimensional $L_p^M$ spaces classical randomized  algorithms are 
essentially as powerful as quantum algorithms.
\end{abstract}

\section{Introduction}
The quantum model of computaton is a theoretical tool to assess 
the potential computing power of quantum mechanics. Quantum algorithms have a random component --
the measurement process. Thus, there is familiarity to classical (that is, non-quantum) randomized algorithms. 
On the other hand,
they also exploit quantum parallelism due to superpositions. (Details and further references can be found, e.g., in
\cite{Hei01a}.) It is a challenging question to compare the computing power of the quantum with the 
classical randomized model. The present paper is devoted to a question of such type.

Mean computation of uniformly bounded scalar sequences is by now well understood in both settings.
In the scalar case quantum computation yields an improvement of the convergence rate by a factor of order $n^{-1/2}$
(where $n$ is the number of function values used) over the randomized
setting. Here we study the vector-valued case.
Our results show that for mean computation in $L_p^M$ spaces of sufficiently high dimension, the speedup of 
quantum over the randomized setting vanishes. 
In particular, there is no vector analogue of the quantum summation algorithm of \cite{BHMT00} 
of comparable efficiency.

Our approach also implies, that the many sums problem, which arose in
\cite{NSW02}, cannot be solved as quickly on a quantum computer as the one sum problem. 
Or in other words, while 
classical algorithms can "re-use" function values for computing further components, 
quantum algorithms fail that property, in general. Details  will be discussed in Section 
\ref{sec:4}.

\section{Notation}

\noindent For $1\le p\le \infty$, $N\in\N$,  and a normed space $X$, 
let 
$$
L_p^N(X) =\{f\,|\,f:\{0,\dots,N-1\}\to X\},
$$ 
equipped with the norm
$$
 \|f\|_{L_p^N(X)}=\left(\frac{1}{N}\sum_{i=0}^{N-1}\|f(i)\|_X^p\right)^{1/p}  
$$
if $1\le p<\infty$, and 
$$
\|f\|_{L_\infty^N(X)}=\max_{0\le i<N}\|f(i)\|_X.
$$
The unit ball is denoted as $\mathcal{B}_p^N(X)=\{f\in L_p^N(X)\,:\,\|f\|\le 1\}$.
Furthermore, we write $L_p^N$ for $L_p^N(\R)$ and $\mathcal{B}_p^N$
for $\mathcal{B}_p^N(\R)$.

We are concerned with the following 
problem: Given $f\in \mathcal{B}_\infty^N(X)$, accessible through function values (at request $i$ 
a subroutine returns $f(i)\in X$), compute (approximately)
$$
S_Nf=\frac{1}{N}\sum_{i=0}^{N-1}f(i)\in X 
$$
for $f\in \mathcal{B}_\infty^N(X).$

We consider the following error settings, which we present in a general context.
 Let $D$ and $K$ be nonempty sets and $F$ 
a nonempty set of $K$-valued functions on $D$. Let $G$ be a normed space and let $S:F\to G$
be a mapping. (In our case $D=\{0,\dots,N-1\}$, $K=X$, $F= \mathcal{B}_\infty^N(X)$, $G=X$, 
and  $S=S_N$.) Let $n\in \N_0$.

In the classical randomized (also called Monte Carlo) setting we define 
\begin{equation}
\label{B2}
e_n^\ran (S,F)=\inf_{A_n}
\sup_{f \in F}\, \E \| S(f) - A_n^\omega(f)\|.
\end{equation}
where the infimum is taken over all $A_n=(A_n^\omega)_{\omega\in \Omega}$ of the form
\begin{equation}   \label{C5}  
A_n^\omega (f) = \phi^\omega (f(t_1^\omega),\ldots, f(t_n^\omega)).
\end{equation}
Here $(\Omega,\Sigma,\Prm)$ is a probability space with $\Omega$ a finite set. For each $\omega\in \Omega $, 
$t_i^\omega\in D\; (1\le i\le n) $ and
$\phi^\omega $ is an arbitrary  mapping from $K ^n$ to $G$. (In the case $n=0$, 
the mappings $\phi^\omega$ are interpreted as not depending on values of $f$, that is, 
as an element of $G$ depending only on $\omega\in \Omega$.)
Consequently, $e_n^\ran (S,F)$ is the best possible error of 
randomized classical (nonadaptive) algorithms using $\le n$
(randomized) function values. (Algorithms with $|\Omega|<\infty$ are usually called restricted randomized
algorithms \cite{Nov88}. For us this setting is technically convenient, and it entails no loss of generality
since we use it only for upper bounds. Comments on lower bounds for the stronger model with arbitray $\Omega$
are given in Section \ref{sec:4}.)

Next let us briefly recall the quantum setting, details and references for further
reading can be found in 
\cite{Hei01}. The notation $\Z[0,M):=\{0,\dots,M-1\}$ will be convenient.
Let $H_1=\C^2$ be the 2-dimensional complex Hilbert space and let 
$\lg 0 \rs$ and $\lg 1 \rs$ be the unit vectors in $H_1$. 
For $m \in \N$ consider the $2^m$-dimensional tensor product space
$$
H_m= \underbrace{H_1 \otimes \dots \otimes H_1}_{m}.
$$
An orthonormal basis is given by the vectors 
$$
\lg \ell  \rs=\lg i_1 \rs\lg i_2 \rs\dots \lg i_m \rs, \qquad \ell\in \Z[0,2^m),
$$
with $\ell = \sum_{j =1}^m i_j \, 2^{m-j}$ 
the binary representation. A quantum query on $F$ is given by a tuple
$$
Q=(m,m',m'', Z, \tau,\b),
$$
where $m,m',m''\in\N$, $m'+m''\le m$,
$Z$ is a nonempty subset of $\Z[0,2^{m'})$, and 
$$
\tau:Z\to D
$$
and 
$$
\b:K\to\Z[0,2^{m''})
$$
are any mappings. 
With a quantum query $Q$ and an input $f\in F$ we associate
a unitary operator $Q_f$ on $H_m$ which is defined 
for $i\in\Z[0,2^{m'})$, $x\in\Z[0,2^{m''})$, $y\in\Z[0,2^{m-m'-m''})$
on the basis state
$$\lg i\rs\lg x\rs\lg y\rs\in
H_{m}=H_{m'}\otimes H_{m''}\otimes H_{m-m'-m''}
$$
as
$$ 
Q_f\lg i\rs\lg x\rs\lg y\rs=
\left\{\begin{array}{ll}
\lg i\rs\lg x\oplus\beta(f(\tau(i)))\rs\lg y\rs &\quad \mbox {if} \quad i\in Z\\
\lg i\rs\lg x\rs\lg y\rs & \quad\mbox{otherwise}\, . 
 \end{array}
\right. 
$$
Here $\oplus$ denotes addition modulo $2^{m''}$.
A quantum algorithm with $n$ quantum queries is a tuple
$$
A_n=(Q,(U_i)_{i=0}^n,b,\varphi),
$$
where $Q$ is a quantum query as defined above,
$U_i$ are any unitary operators on $H_m$, $b\in \Z[0,2^m)$, and
$$
\varphi:\Z[0,2^m)\to G
$$
is an arbitrary mapping. 
At input $f\in F$ the algorithm produces
the state
\begin{equation}
\label{FA1}
\lg \psi\rs=U_n Q_f U_{n-1}\ldots U_1 Q_f U_0 \lg b\rs.
\end{equation}
(The symbol $\lg \psi\rs$ means any element of the unit sphere 
of $H_m$, while symbols like $\lg i\rs, \lg x\rs, \lg l\rs$ with $i,x,l$ integers
stand for basis states.) Let
$$      
\lg \psi\rs=\sum_{\ell=0}^{2^m-1} \alpha_{\ell,f} \lg \ell\rs \,  .
$$
The state $\lg \psi\rs$ is measured, which means that we obtain a
random variable $\xi_f(\omega)$ with values in
$\{0,\ldots ,2^m-1\}$, which takes the value $\ell$ with probability
$|\alpha_{\ell,f}|^2$. 
Finally the mapping $\phi$ is applied, so the output of the algorithm is 
\begin{equation}
\label{C6}
A_n(f,\omega)=\phi(\xi_f(\omega)).
\end{equation}
What we described here is a quantum algorithm with one measurement. Such algorithms 
turn out to be 
essentially equivalent to algorithms with several measurements (see \cite{Hei01}, 
Lemma 1).
Now we define
\begin{equation}
\label{B3}
e_n^\q (S,F)=
\inf_{A_n} \sup_{f\in F}\inf\left\{\varepsilon:
\,\Prm\{|S(f)-A_n(f,\omega)|\le\varepsilon\}\ge 3/4 \right\}.
\end{equation}
The first infimum is taken over all quantum algorithms $A_n$ which use not more than 
$n$ quantum queries. 
Thus, $e_n^\q (S,F)$ is the best possible error of quantum 
algorithms using $\le n$
quantum queries. 

\section{Mean Computation}

In the case of $X=\R$, the rates are known in both settings. We summarize these previous results in 
\begin{theorem}\label{theo:1}
There are constants $c_0,c_1,c_2>0$ such that for $n,N\in \N$ with $n\le c_0N$
\begin{eqnarray}
c_1 n^{-1/2} &\le& e_n^\ran (S_N,\mathcal{B}_\infty^N)\le c_2 n^{-1/2}\label{A2}\\
c_1 n^{-1} &\le& e_n^\q(S_N,\mathcal{B}_\infty^N) \;\;\le  c_2n^{-1}.\label{A3}
\end{eqnarray}
\end{theorem}
We often use the 
same symbol $c,c_1,\dots$ for possibly different
positive constants -- also when they appear in a sequence of relations.
The estimate (\ref{A2}) is well-known, see e.g.\ \cite{Nov88}, \cite{TWW88}, \cite{M95},
while the upper bound of (\ref{A3})  is due to Brassard, H{\o}yer, 
Mos\-ca, and Tapp \cite{BHMT00}, the lower bound to
Nayak and Wu \cite{NW}, see also \cite{Hei01}.

Considering the vector valued case, two natural problems arise: 
\begin{itemize}
\item Does the speedup of the quantum over the randomized setting extend to the vector valued case?
\item Can the many sums problem be solved as quickly on a quantum computer 
as the  one sum problem?
\end{itemize}
Interest in the many sums problem arose in \cite{NSW02}. It is the following problem:
Let $M\in \N$ and $X=\R^M$, 
equipped with any norm.
We are given
a (completely known to us)  $a\in L_\infty^N(X)$, that is, a sequence of vectors
$$
a(j)=(a_{ij})_{i=0}^{M-1}\in X\quad (j=0,\dots,N-1),
$$
the task is, given $f\in L_\infty^N$, accessible through function values,
compute 
$$
T_N^af=\frac{1}{N}\sum_{j=0}^{N-1}f(j)a(j)
=\left(\frac{1}{N}\sum_{j=0}^{N-1}a_{ij}f(j)\right)_{i=0}^{M-1}
\in X,
$$ 
or, in other words, compute the (weighted) means for $i=0,\dots,M-1$ simultaneously, the error being measured in the
norm of $X$.

First we show how the many sums problem can be reduced to vector-valued mean computation. 
\begin{lemma}\label{lem:1}
For all $n\in \N_0$, $N\in \N$,
$$ 
\sup_{a\in \mathcal{B}_\infty^N(X)} e_n^\ran(T_N^a,\mathcal{B}_\infty^N)\le e_n^\ran(S_N,\mathcal{B}_\infty^N(X)),
$$
and
$$ 
\sup_{a\in \mathcal{B}_\infty^N(X)} e_{2n}^\q(T_N^a,\mathcal{B}_\infty^N)\le e_n^\q(S_N,\mathcal{B}_\infty^N(X)).
$$
\end{lemma}
\begin{proof}
Let $a\in \mathcal{B}_\infty^N(X)$ and define
$$
V^a: \mathcal{B}_\infty^N\to \mathcal{B}_\infty^N(X)
$$
by setting for $f\in\mathcal{B}_\infty^N$ and $0\le j <N$
$$
(V^af)(j)= f(j)a(j)\in X.
$$
Then $T_N^a=S_NV^a$, and
the result for $e_n^\ran$ follows immediately from the definition.

In the quantum case we need some technicalities connected with 
the special finite representation in the form of the query.
Let $m^*\in \N$  be arbitrary and define
$\b:\R\to\Z[0,2^{m^*})$ for $z\in\R$ by
\begin{equation}\label{N1}
\b(z)=
\left\{\begin{array}{lll}
   0& \mbox{if} \quad z <-1 \\
   \lfloor 2^{m^*-1}(z+1)\rfloor       & \mbox{if} \quad  
   -1\le z <1\\
   2^{m^*}-1& \mbox{if} \quad z\ge 1. 
   \end{array}
   \right.
\end{equation}
Furthermore, let $\gamma:\Z[0,2^{m^*})\to\R$ be defined for $y\in\Z[0,2^{m^*})$ as 
\begin{equation}
\label{N2}
\g(y)=2^{-m^*+1}y-1.
\end{equation}
It follows that for $-1\le z\le 1$,
\begin{equation}
\label{E4}
-1\le\g(\b(z))\le z\le \g(\b(z))+2^{-m^*+1}\le 1.
\end{equation}
Define $\Gamma: \mathcal{B}_\infty^N\to \mathcal{B}_\infty^N(X)$ by
\begin{equation}
\label{A4}
(\Gamma f)(i)=(\g\circ\b\circ f)(i)\, a(i).
\end{equation}
This mapping is of the form (13) of \cite{Hei01} with $\kappa=1$, 
$\eta:D\to D$ the identity on $D=\Z[0,N)$, $\b$ as above, and 
$\rho:D\times\Z[0,2^{m^*})\to X$ being defined as
$$
\rho(i,z)=\g(z)a(i).
$$
Hence, by Corollary 1 of \cite{Hei01},
\begin{equation}
\label{A6}
e_{2n}^\q(S_N\Gamma,\mathcal{B}_\infty^N)\le e_{n}^\q(S_N,\mathcal{B}_\infty^N(X)).
\end{equation}
By (\ref{A4}), 
$$
 S_N\Gamma f= S_N V^a (\g\circ\b\circ f)=T_N^a(\g\circ\b\circ f).
$$
Because of (\ref{E4}) we have 
\begin{eqnarray*}
\sup_{f\in \mathcal{B}_\infty^N}\|T_N^a f- S_N\Gamma f\|&=&
\sup_{f\in \mathcal{B}_\infty^N}\|T_N^a f-T_N^a(\g\circ\b\circ f)\|\\
&\le &2^{-m^*+1}\|T_N^a\|
\le2^{-m^*+1}.
\end{eqnarray*}
This together with Lemma 6 (i) of \cite{Hei01} and  (\ref{A6}) above implies
\begin{eqnarray*}
e_{2n}^\q(T_N^a,\mathcal{B}_\infty^N)&\le& 
e_{2n}^\q( S_N\Gamma,\mathcal{B}_\infty^N)+2^{-m^*+1}\\
&\le& e_{n}^\q(S_N,\mathcal{B}_\infty^N(X))+2^{-m^*+1}.
\end{eqnarray*}
Since $m^*$ can be made arbitrarily large, the desired result follows.
\end{proof}

To derive upper bounds for the randomized setting, we need some concepts and results from
the theory of Banach space valued random variables, which can be found in \cite{LT91}.
Let $1<p\le 2$. For a Banach space $X$ the type $p$ constant of $X$ is 
the smallest $c>0$ such that
 for all $m$ and all $x_1, \dots , x_m \in X$
\begin{eqnarray*}
\E \Vert \sum_{i=1}^m \varepsilon_i x_i \Vert_X^p \le c^p \sum_{i=1}^m \Vert x_i \Vert_X^p,
\end{eqnarray*}
where $(\varepsilon_i)_{i=1}^m$ are independent, centered, $\{-1,1\}$ valued Bernoulli 
variables. We put $T_p(X)=+\infty$ if there is no such $c$.
It is known that for $1\le p\le \infty$ there is a constant $c>0$ such that 
for all $M\in \N$
\begin{eqnarray}
T_p(L_p^M)&\le& c\quad \mbox{if}\quad 1<p<2  \label{C1}\\
T_2(L_p^M)&\le& c\quad \mbox{if}\quad 2\le p<\infty\label{C2}\\
T_2(L_\infty^M)&\le& c\, (\log (M+1))^{1/2}  \label{C3}.
\end{eqnarray}
(Throughout this paper $\log$ stands for $\log_2$.)
Using the concepts above, the upper bounds for the classical randomized setting, first observed 
by Math\'e \cite{M94}, are easily derived:
\begin{proposition}\label{pro:2}
Let $1\le p \le 2$. Then for 
 $n\in \N$ and any Banach space $X$,
$$
e_n^{\ran}(S_N,\mathcal{B}_\infty^N(X))\le 4\, T_p(X)n^{1/p-1}.
$$ 
\end{proposition}
\begin{proof}
We use the vector valued Monte Carlo method
$$
S_Nf\approx A_n(f) =\frac{1}{n}\sum_{l=1}^n f(\xi_l),
$$ 
where $\xi_l$ are uniformly distributed on $\{0,\dots,N-1\}$  
independent random variables. By Proposition 9.11 of \cite{LT91} 
\begin{eqnarray*}
\E\|S_Nf- A_n(f)\|_{X}
&\le&(\E\|S_Nf- A_n(f)\|_{X}^p)^{1/p}\\
&=&n^{-1}(\E\|\sum_{l=1}^n (S_Nf-f(\xi_l))\|_{X}^p)^{1/p}\\
&\le &2T_p(X)n^{-1} (\sum_{l=1}^n\E\|S_Nf-f(\xi_l))\|_{X}^p)^{1/p}\\
&\le &4T_p(X)n^{1/p-1}\max_{0\le i<N}\|f(i)\|_X.
\end{eqnarray*}
\end{proof}
Next we show that these upper bounds can be carried over to 
the quantum setting. Indeed, we shall show that one can 
transform any randomized algorithm $A_n^\ran$  into a quantum algorithm $A_n^\q$ with $n$ queries and 
essentially the same error behaviour. We consider this in the general context in
which we defined the error quantities.
\begin{lemma}
\label{lem:2} Let $n\in \N_0$, let  
$A_n^\ran=(A_n^\omega)_{\omega\in\Omega}$ 
be any randomized algorithm on F using $n$ function values (recall that we assume a finite
$\Omega $). Suppose further that $\theta: K\to K$ is a mapping such that
$\theta(K)$ is a finite set and for all $f\in F$ we have $\theta\circ f\in F$. Then 
there is a quantum algorithm $A_n^\q$ using $n$ queries such that for all $f\in F$ the
distribution of $A_n^\q(f)$ is the same as that of $A_n^\ran(\theta\circ f)$. Moreover,
\begin{equation}
\label{D1}
e_n^\q(S,F)\le 4 e_n^\ran(S,F)+\sup_{f\in F}\|S(f)-S(\theta\circ f)\|.
\end{equation}
\end{lemma}
\begin{proof} Choose any $m'_1,m'_2,m''\in \N$ satisfying
$$
n\le 2^{m'_1},\quad |\Omega|\le  2^{m'_2},\quad |\theta(K)|\le 2^{m''}.
$$
Put $m'=m'_1+m'_2$ and $m=m'+nm''$. 
It is no loss of generality to assume 
$\Omega\subseteq\Z[0,2^{m'_2})$ 
and $\{\omega\}\in\Sigma $ for all $\omega\in \Omega$.

Let $\sigma$
be any bijection of $\theta(K)$ onto a subset $Z_0$ of $\Z[0,2^{m''})$. Setting
$\b=\sigma\circ\theta: K\to \Z[0,2^{m''})$ 
and letting $\gamma:\Z[0,2^{m''})\to K$ be
any extension of $\sigma^{-1}:Z_0\to \theta(K)$ to all of $\Z[0,2^{m''})$, we 
get%
\begin{equation}
\label{G1}
\g\circ\b=\theta.
\end{equation}
 We identify $\Z[0,2^{m'})$ with 
$\Z[0,2^{m'_1})\times \Z[0,2^{m'_2})$.
Let
$$
Z=\{(i,\omega): \;0 \le i \le n-1,\,\omega\in \Omega\}\subseteq \Z[0,2^{m'}),
$$ 
and define $\tau:Z\to D$ as $\tau(i,\omega)=t_{i+1}^\omega$.
The query of $A_n^\q$ is given by  
$$
Q=(m,m',m'', Z, \tau,\b), 
$$
The quantum algorithm $A_n^\q$ is defined by
$$
A_n^\q=(Q,(U_i)_{i=0}^n, 0,\varphi),
$$
where $U_i$ and $\varphi$ will be specified in the sequel. 
As $U_0$ we choose any unitary operator on $H_m$ mapping 
$$
\lg 0\rs\lg 0\rs\lg 0\rs\in H_{m'_1}\otimes H_{m'_2}\otimes H_{m-m'}=H_m
$$
to 
\begin{equation}
\label{CB3}
\sum_{\omega\in\Omega} \Prm(\{\omega\})^{1/2}\lg n-1\rs\lg \omega\rs\lg 0\rs.
\end{equation}
(In the case $n=0$ we replace $n-1$ by 1.) Applying $Q_f$ maps a state of the form  
$$
\lg n-1\rs\lg \omega\rs\lg 0\rs\dots\lg 0\rs\lg 0\rs
\in H_{m'_1}\otimes H_{m'_2}\otimes \underbrace{H_{m''}\otimes\dots\otimes H_{m''}\otimes H_{m''}}_n =H_m
$$
with $\omega\in \Omega$ to 
$$
\lg n-1\rs\lg \omega\rs\lg \beta(f(t_n^\omega))\rs\dots\lg 0\rs\lg 0\rs.
$$
Now $U_1$ is any unitary mapping that decreases the first component by one 
and interchanges the third with the last, that is, maps the state above to
$$
\lg n-2\rs\lg \omega\rs\lg 0\rs\dots\lg 0\rs\lg \beta(f(t_n^\omega))\rs
$$
(for all $\omega\in \Omega$). The next application of $Q_f$, followed by a $U_2$ 
which decreases the first component by one and interchanges the third with the last but one, 
results in
$$
\lg n-3\rs\lg \omega\rs\lg 0\rs\dots\lg \beta(f(t_{n-1}^\omega))\rs
\lg \beta(f(t_n^\omega))\rs.
$$
Continuing this way we reach, after the $n$-th application of 
$Q_f$, the state 
$$
\lg 0\rs\lg \omega\rs\lg \beta(f(t_1^\omega))\rs
\dots\lg \beta(f(t_{n-1}^\omega))\rs\lg \beta(f(t_n^\omega))\rs.
$$ 
The last $U_n$ does not change anything, that is, it is the identity on $H_m$.
Finally, $\phi:\Z[0,2^m)\to G$ is defined for
\begin{eqnarray*}
\lefteqn{   
 (i,\omega,z_1,\dots,z_n) }\\
&\in &\Z[0,2^{m'_1})\times\Z[0,2^{m'_2})
\times\Z[0,2^{m''})\times\dots\times\Z[0,2^{m''})=\Z[0,2^m)
\end{eqnarray*}
by
$$
\phi(i,\omega,z_1,\dots,z_n)=\left\{\begin{array}{lll}
  \phi^\omega(\g(z_1),\dots,\g(z_n))\in G & \mbox{if} \quad \omega \in \Omega   \\
   0\in G & \mbox{otherwise.}    \\
    \end{array}
\right. 
$$ 
Taking into account (\ref{G1}) and (\ref{CB3}), it follows readily that $A_n^\q(f)$ takes the value
$$
\phi^\omega(\g\circ\beta\circ f(t_1^\omega),\dots,
\g\circ\beta\circ f(t_n^\omega))
=\phi^\omega(\theta\circ f(t_1^\omega),\dots,
\theta\circ f(t_n^\omega))
$$
with probability $|\Prm(\{\omega\})^{1/2}|^2=\Prm(\{\omega\})$. So does 
$A_n^\ran(\theta\circ f)$, by definition, which proves the first 
statement of the lemma.

To show (\ref{D1}), we fix any $\delta>0$ and assume that $A_n^\ran$ satisfies
$$
\sup_{f \in F}\, \E \| S(f) - A_n^\omega(f)\|\le  
e_n^\ran (S,F) +\delta.
$$
By Chebyshev's inequality, for any $f\in F$,
\begin{equation*}
\Prm\{\omega:\;\|S(\theta\circ f)-A_n^\omega(\theta\circ f)\|
\le 4\, \E\|S(\theta\circ f)-A_n^\omega(\theta\circ f)\|\}\ge 3/4.
\end{equation*}
Hence, with probability at least 3/4, we also have 
$$
\|S(\theta\circ f)-A_n^\q(f)\|\le 4 
\E\|S(\theta\circ f)-A_n^\omega(\theta\circ f)\|
\le 4e^\ran(S,F)+4\delta,
$$
and consequently,
\begin{eqnarray*}
\|S(f)-A_n^\q(f)\|&\le &\|S(f)-S(\theta\circ f)\|+\|S(\theta\circ f)-A_n^\q(f)\|\\
&\le &\sup_{f\in F}\|S(f)-S(\theta\circ f)\|+4 e_n^\ran(S,F)+4\delta.
\end{eqnarray*}
Since $\delta>0$ was arbitrary, this proves (\ref{D1}).
\end{proof}
\begin{corollary}\label{cor:2}
If $n\in \N_0$, $N\in \N$, $F=\mathcal{B}_\infty^N(X)$ with $X$ a finite dimensional normed space, and
$S:F\to G$ is any continuous mapping to a normed space G, then 
$$
e_n^\q(S,F)\le 4 e_n^\ran(S,F).
$$
\end{corollary}
\begin{proof}
Since $X$ is finite dimensional, the unit ball $B_X$ is compact. Using finite $(1/k)$-nets in $B_X$,
we can choose $\theta_k:X\to B_X \;(k\in \N)$ in such a way that $\theta_k(X)$ is finite for all $k$ and
$$
\lim_{k\to\infty}\sup_{x\in B_X} \|x-\theta_k(x)\|_X =0.
$$
This implies 
$$
\lim_{k\to\infty}\sup_{f\in \mathcal{B}_\infty^N(X)}\|f -\theta_k\circ f\|_{L_\infty^N(X)}=0.
$$
The set $F=\mathcal{B}_\infty^N(X)$ is also compact. Therefore, $S$ is uniformly continuous on $F$.
It follows that 
$$
\lim_{k\to\infty}\sup_{f\in F}\|S(f) -S(\theta_k\circ f)\|_G=0,
$$
and the result is a consequence of Lemma \ref{lem:2}.
\end{proof}
Taking as $X$ finite dimensional $L_p$-spaces, we get from Proposition \ref{pro:2}, 
 (\ref{C1})--(\ref{C3}), and Corollary \ref{cor:2},
\begin{corollary}\label{cor:3}
 Let $1\le p \le\infty$. Then there is a constant $c>0$ such that for 
 $n,M,N\in \N$, and $X=L^M_p$ 
\begin{eqnarray*}  
\lefteqn{e_n^\q (S_N,\mathcal{B}_\infty^N(X))\le 4 e_n^\ran (S_N,\mathcal{B}_\infty^N(X))}\\[.1cm]
&\le&
c\left\{ \begin{array}{ll}
n^{-1/2}&\quad \mbox{if}\quad 2\le p<\infty\\
n^{-1/2}(\log (M+1))^{1/2}&\quad \mbox{if}\quad p =\infty\\
n^{-1+1/p}&\quad \mbox{if}\quad 1\le p<2. \\
\end{array} 
\right. 
\end{eqnarray*}
\end{corollary}
To derive lower bounds, we need two simple technical lemmas. 
\begin{lemma}\label{lem:3}
For all $n\in \N_0$, $N,N_1\in \N$ with $N_1\le N$ and  
normed spaces $X$ the following holds: Let 
$N=kN_1+l$ for some $k\in \N$, $l\in \N_0$ with $0\le l<N_1$. Then
$$
\sup_{a\in \mathcal{B}_\infty^{N_1}(X)} e_n^\q(T^a_{N_1},\mathcal{B}_\infty^{N_1})\le 
\frac{N}{kN_1}\sup_{a\in \mathcal{B}_\infty^{N}(X)} e_n^\q(T^a_{N},\mathcal{B}_\infty^{N})
<2\sup_{a\in \mathcal{B}_\infty^{N}(X)} e_n^\q(T^a_{N},\mathcal{B}_\infty^{N}).
$$
\end{lemma}
\begin{proof} The second inequality is obvious. To prove the first, let 
$a\in \mathcal{B}_\infty^{N_1}(X)$. Define $\tilde{a}\in \mathcal{B}_\infty^{N}(X)$
by 
$$
\tilde{a}(i)=\left\{\begin{array}{lll}
  a(i \bmod N_1)  & \mbox{if} \quad i<kN_1    \\
  0 & \mbox{if}   \quad kN_1\le i<N.  \\
    \end{array}
\right.
$$ 
Define $\Gamma:\mathcal{B}_\infty^{N_1}\to \mathcal{B}_\infty^{N}$ by setting 
for $f\in L_\infty^{N_1}$ 
\[
(\Gamma f)(i)=
  f(i \bmod N_1). 
\]
Then $\Gamma$ is of the form (12) of \cite{Hei01}. For 
$f\in \mathcal{B}_\infty^{N_1}$,
$$
T_N^{\tilde{a}}\Gamma f=\frac{1}{N}\sum_{i=0}^{N-1}(\Gamma f)(i)\tilde{a}(i)
=\frac{1}{N}\sum_{j=0}^{N_1-1} kf(j)a(j)=\frac{kN_1}{N} T^a_{N_1}f.
$$
By \cite{Hei01}, Corollary 1 and Lemma 6 (ii), the result follows.
\end{proof}
\begin{lemma}\label{lem:4}
For all $n\in \N_0$, $N,M,M_1\in \N$ with $M_1\le M$, and $1\le p\le \infty$,
$$
\sup_{a\in\mathcal{B}_\infty^{N}(L_p^{M_1})}e_n^\q(T^a_{N},\mathcal{B}_\infty^{N})\le 
\sup_{a\in\mathcal{B}_\infty^{N}(L_p^{M})}e_n^\q(T^a_{N},\mathcal{B}_\infty^{N}).
$$
\end{lemma}
\begin{proof}
Define $J:L_p^{M_1}\to L_p^M$ by setting for $g\in L_p^{M_1}$ 
\[
(Jg)(i)=\left\{\begin{array}{cll}
  \left(\frac{M}{M_1}\right)^{\frac{1}{p}} g(i) & \mbox{if} \quad i<M_1   \\
  0 & \mbox{if} \quad M_1\le i <M,  \\
    \end{array}
\right. 
\]
and $P:L_p^M\to L_p^{M_1}$ for $g\in L_p^{M}$  by
\[
(Pg)(i)=
  \left(\frac{M_1}{M}\right)^{\frac{1}{p}} g(i) \quad  (0\le i<M_1).
\]
Clearly, 
$$
\|P\|=\|J\|=1.
$$
Let $a\in \mathcal{B}_\infty^{N}(L_p^{M_1})$. 
Define $\tilde{a}\in \mathcal{B}_\infty^{N}(L_p^{M})$ by  
\[
\tilde{a}(j)=Ja(j)\quad (0\le j<N).
\]
Then for $f\in  \mathcal{B}_\infty^{N}$
$$
PT_N^{\tilde{a}} f=\frac{1}{N}\sum_{j=0}^{N-1} f(j)P\tilde{a}(j)
=\frac{1}{N}\sum_{j=0}^{N-1} f(j)a(j)=T_N^af.
$$
By Lemma 1 of \cite{Hei03a},
$$
e_n^\q(T^a_{N},\mathcal{B}_\infty^{N})=e_n^\q(PT_N^{\tilde{a}},\mathcal{B}_\infty^{N})
\le \|P\|e_n^\q(T_N^{\tilde{a}},\mathcal{B}_\infty^{N})= e_n^\q(T_N^{\tilde{a}},\mathcal{B}_\infty^{N}).
$$
\end{proof}
We also need the following result which is contained in Proposition 6 of \cite{Hei03a}. 
Here $J_{\infty,p}^{N}:L_\infty^N\to L_p^N$
denotes the identical embedding. For brevity we set for $N\in \N$ with $N>4$
$$
\lambda(N):=(\log\log N)^{-3/2}(\log\log\log N)^{-1}.
$$ 
\begin{proposition}
\label{pro:7} 
Let $1\le p\le  \infty$. There are constants $c_0,c_1>0$ such that 
for all $n,N\in\N$ with $N>4$ and $n\le c_0 N$ 
$$
e_n^\q(J_{\infty,p}^{N},\mathcal{B}_\infty^N)\ge c_1
\left\{\begin{array}{lll}
 1  & \mbox{if} \quad 2<p\le\infty,   \\
 \lambda(N)  & \mbox{if} \quad  p=2,  \\ 
  (\log N)^{-2/p+1}   & \mbox{if} \quad 1\le p<2.   \\
    \end{array}
\right. 
$$
\end{proposition}
Note that, necessarily, $c_0<1$, since $e_N^\q(J_{\infty,p}^{N},\mathcal{B}_\infty^N)=0$ (this is easy to check, 
for an argument of this type see \cite{HN01b}, relation (12) and its proof). 
\begin{proposition}\label{pro:3}
 Let $1\le p \le\infty$. Then there are constants $c_0,c_1>0$ such that for 
 $n,M,N\in \N$ with $4<n\le c_0\min(M,N)$, and $X=L^M_p$
$$
\sup_{a\in \mathcal{B}_\infty^{N}(X)} e_n^\q(T^a_{N},\mathcal{B}_\infty^{N})
\ge c_1
\left\{ \begin{array}{ll}
n^{-1/2}\lambda(n)&\quad \mbox{if}\quad 2\le p\le\infty\\
n^{-1+1/p}(\log n)^{-2/p+1}&\quad \mbox{if}\quad 1\le p<2. \\
\end{array} 
\right. 
$$
\end{proposition}
\begin{proof}
Assume
\begin{equation}
\label{CF1}
4<n\le (c_0/2)\min(M,N),
\end{equation}
where the constant $c_0$ is that from Proposition \ref{pro:7}.
Define 
$$
k=\lceil\log_2(c_0^{-1}n)\rceil, \quad N_1=2^k.
$$
It follows that 
$$
c_0^{-1}n\le N_1\le 2c_0^{-1}n,
$$
which implies $n\le c_0N_1$, and, because of $c_0<1$ and (\ref{CF1}), also 
$$
4<N_1\le \min(M,N).
$$
From this and Lemmas \ref{lem:3} and \ref{lem:4} we obtain
\begin{eqnarray}
\label{E1}
\sup_{a\in \mathcal{B}_\infty^{N_1}(L_p^{N_1})} e_n^\q(T^a_{N_1},\mathcal{B}_\infty^{N_1})&\le&
2\sup_{a\in \mathcal{B}_\infty^{N}(L_p^{N_1})} e_n^\q(T^a_{N},\mathcal{B}_\infty^{N})\nonumber\\
&\le&
2\sup_{a\in \mathcal{B}_\infty^{N}(L_p^{N})} e_n^\q(T^a_{N},\mathcal{B}_\infty^{N}).
\end{eqnarray}
For $1\le p< 2$ define $a\in \mathcal{B}_\infty^{N_1}(L_p^{N_1})$ by  
$$
a(j)=N_1^{1/p}e_j\quad(0\le j < N_1-1),
$$ 
where  $e_j=(\delta_{ij})_{i=0}^{N_1-1}$ are 
the unit vectors in $\R^{N_1}$.  Then
$$
T_{N_1}^a=N_1^{1/p-1}J_{\infty,p}^{N_1}.
$$ 
Thus, from Proposition \ref{pro:7}
$$
e_{n}^\q (T_{N_1}^a,\mathcal{B}_\infty^{N_1})\ge c_1 N_1^{1/p-1}(\log N_1)^{-2/p+1}
\ge  c n^{1/p-1}(\log n)^{-2/p+1},
$$
which, together with (\ref{E1}) implies the required statement. 

For $2\le p\le \infty$ we recall that $N_1=2^k$ and let  
$W_{N_1}$ be the Walsh matrix, defined by 
$$
W_{N_1}=\left((-1)^{i \cdot j}\right)_{i,j=0}^{N_1-1}.
$$ 
Here
$$ 
i \cdot j:=\sum_{l=1}^k i_l j_l
$$
with $i=\sum_{l =1}^k i_l \, 2^{k-l}$ and $j=\sum_{l =1}^k j_l \, 2^{k-l}$ the
binary representations. Note that 
\begin{equation}
\label{CB1}
W_{N_1}^2=N_1I_{N_1},
\end{equation}
where $I_{N_1}$ is the respective
identity matrix.
Let $w(j)$ denote the $j$-th column vector of $W_{N_1}$ and define 
$a\in \mathcal{B}_\infty^{N_1}(L_p^{N_1})$ by  
$$
a(j)=w(j) \quad (0\le j<N_1-1).
$$ 
Let $W$ denote the operator from $L_p^{N_1}$ to $L_2^{N_1}$
with matrix $W_{N_1}$, that is,
$$
Wf=\sum_{j=0}^{N_1-1}f(j)w(j)\quad (f\in L_p^{N_1}).
$$
It follows from (\ref{CB1}) that for $f\in L_\infty^{N_1}$,
$$
WT_{N_1}^af=\frac{1}{N_1}W\sum_{i=0}^{N_1-1}f(j)w(j)= \sum_{i=0}^{N_1-1}f(j)e_j,
$$
and consequently
$$
WT_{N_1}^a=J_{\infty,2}^{N_1}.
$$
On the other hand, since $W_{N_1}$ is an orthogonal matrix, 
\begin{eqnarray*}
\|W:L_p^{N_1}\to L_2^{N_1}\|
&\le& \|I_{N_1}:L_p^{N_1}\to L_2^{N_1}\|\|W:L_2^{N_1}\to L_2^{N_1}\|\\
&\le& \|W:L_2^{N_1}\to L_2^{N_1}\|=N_1^{1/2}.
\end{eqnarray*}
It follows from Lemma 1 of \cite{Hei03a} that
$$
e_{n}^\q(J_{\infty,2}^{N_1},\mathcal{B}_\infty^{N_1})
=e_{n}^\q(WT_{N_1}^a,\mathcal{B}_\infty^{N_1})\le \|W
\|e_{n}^\q(T_{N_1}^a,\mathcal{B}_\infty^{N_1})
\le N_1^{1/2}e_{n}^\q(T_{N_1}^a,\mathcal{B}_\infty^{N_1}).
$$
Hence, by Proposition \ref{pro:7},
$$
e_{n}^\q (T_{N_1}^a,\mathcal{B}_\infty^{N_1})
\ge c_1 N_1^{-1/2}\lambda(N_1)
\ge c n^{-1/2}\lambda(n).
$$
Now the result follows from (\ref{E1}).
\end{proof}
From Proposition \ref{pro:3},  Lemma \ref{lem:1}, and Corollary \ref{cor:2} we get
\begin{corollary}\label{cor:4}
 Let $1\le p \le\infty$. Then there are constants $c_0,c_1>0$ such that for 
 $n,M,N\in \N$ with $4<n\le c_0\min(M,N)$, and $X=L^M_p$
\begin{eqnarray*}  
\lefteqn{e_n^\ran(S_{N},\mathcal{B}_\infty^{N}(X))\ge 4^{-1}e_n^\q(S_{N},\mathcal{B}_\infty^{N}(X))
}\\
&\ge& c_1
\left\{ \begin{array}{ll}
n^{-1/2}\lambda(n)&\quad \mbox{if}\quad 2\le p\le\infty\\
n^{-1+1/p}(\log n)^{-2/p+1}&\quad \mbox{if}\quad 1\le p<2. \\
\end{array} 
\right. 
\end{eqnarray*}
\end{corollary}
We summarize the main results (contained in Corollaries \ref{cor:3} and  \ref{cor:4}) 
in the following theorem, in which 
we suppress logarithmic factors. 
\begin{theorem}\label{theo:2}
 Let $1\le p \le\infty$. Then there is a constant $c_0$ such that for 
 $n,M,N\in \N$ with $n\le c_0\min(M,N)$, and $X=L^M_p$
\begin{eqnarray*} 
e_n^\ran(S_N,\mathcal{B}_\infty^N(X))
\asymp_{\log}e_n^\q (S_N,\mathcal{B}_\infty^N(X))
\asymp_{\log}\left\{ \begin{array}{ll}
n^{-1/2}&\quad \mbox{if}\quad 2\le p\le\infty\\
n^{-1+1/p}&\quad \mbox{if}\quad 1\le p<2. \\
\end{array} 
\right. 
\end{eqnarray*}
\end{theorem}
\section{Comments}
\label{sec:4}
In the scalar case, by Theorem \ref{theo:1}, there is a quantum algorithm with rate $n^{-1}$, while no 
classical randomized algorithm can be better than of the order $n^{-1/2}$. Theorem \ref{theo:2}
shows that vector quantum summation fails to give any speedup over 
the classical randomized setting
for any $1\le p\le \infty$, provided the dimension is high ($M\ge N$). Under this assumption there is, in particular,
no vector-valued version of the quantum summation algorithm of \cite{BHMT00} which is more efficient than the 
classical vector-valued Monte Carlo algorithm.

Let us also discuss the many sums problem. 
This is best done for the case $p=\infty$, that is, $X=L_\infty^M$. Let $a\in\mathcal{B}_\infty^N(L_\infty^M)$, thus
$|a_{ij}|\le 1$ for all $i,j$. We can compute one component of the result,
$$
\left(\sum_{j=0}^{N-1} a_{ij}f(j)\right)_{i=0}^{M-1}
$$
say, the first, by the scalar-valued Monte Carlo method:
$$
\frac{1}{n}\sum_{l=1}^na_{1,\xi_l} f(\xi_l),
$$ 
where $\xi_l$ are uniformly distributed on $\{0,\dots,N-1\}$  
independent random variables. This has error rate $n^{-1/2}$. 
Once the function 
values $f(\xi_l)$ have been obtained, they can be re-used in the computation of all other components of the solution vector.
All we have to take care is that the probability of having
the desired precision in all components is large enough -- this way we just lose a logarithmic factor 
(see Corollary \ref{cor:3}).
Now, can we do so in the quantum setting, that is, can we re-use query results? If so,  we should be able to
obtain the same rate $n^{-1}$ (maybe, again, up to a logarithmic factor) as in the scalar case.
It turned out that this is not the case, that is, there are matrices 
$a\in\mathcal{B}_\infty^N(L_\infty^M)$ such that
the best rate is (up to logarithms) $n^{-1/2}$ (Proposition \ref{pro:3}).
  
Let us finally mention that the lower bounds for the randomized setting (Corollary \ref{cor:4}),
obtained here as a byproduct of the analysis of the quantum case, can be slightly improved and
extended to the case of randomized algorithms with general (i.e., also infinite) $\Omega$. Let us denote the respective
minimal error by $\bar{e}_n^{\,\ran}$ (trivially, $e_n^\ran \ge \bar{e}_n^{\,\ran}$). The following holds for $1\le p \le\infty$:
 There are constants $c_0,c_1>0$ such that for 
 $n,M,N\in \N$ with $n\le c_0\min(M,N)$, and $X=L^M_p$
\begin{eqnarray*}  
\bar{e}_n^{\,\ran} (S_N,\mathcal{B}_\infty^N(X))\ge c_1
\left\{ \begin{array}{ll}
n^{-1/2}&\quad \mbox{if}\quad 2\le p<\infty\\
\min(n^{-1/2}(\log (M+1))^{1/2}, 1)&\quad \mbox{if}\quad p =\infty\\
n^{-1+1/p}&\quad \mbox{if}\quad 1\le p<2. \\
\end{array} 
\right. 
\end{eqnarray*}
For $2\le p<\infty$ this follows from the 
scalar case $X=\R$ (Theorem \ref{theo:1}). The case $1\le p<2$ is a direct consequence
of standard lower bound techniques \cite{Nov88}, \cite{TWW88}, \cite{Hei93}. So is the case $p=\infty$, 
except that, in addition,
Lemma 5.3 of \cite{HS99} has to be used.
We omit further details.

\end{document}